
\magnification=\magstep1
\settabs 18 \columns
\baselineskip=17 pt

\def\b{\bigskip}
\def\bb{\bigskip\bigskip}
\def\bbb{\bigskip\bigskip\bigskip}

\def\mybox{\sqcap\kern-.66em\sqcup\kern.66em}

\def\no{\noindent}

\def\ce{\centerline}
\def\ve{\vfill\eject}

\def\e e{$e^+ e^-$ }

\rightline {hep-th/9306048}
\rightline {UCLA/93/TEP/17}
\rightline{NYU-TH.93/06/01}
\rightline {June 1993}
\bbb
\ce {\bf STRING PHASE TRANSITIONS IN A STRONG MAGNETIC FIELD\footnote{*}
{\rm{Work
suported in part by the United States Department of Energy under Contract No.
DOE-AT03-88ER40384, Task C.}}}
\bb
\ce {\bf S. Ferrara}

\b
\ce {\it Theory Division, CERN, 1211 Geneva 23, Switzerland}
\b
\ce {and}
\b
\ce {\bf M. Porrati\footnote{**}{\rm{On leave of absence from INFN, Sezione di
Pisa, Italy.}}}
\b
\ce {\it Department of Physics, New York University, New York, NY 10003}
\bb
\ce {\bf Abstract}
\b
\no We consider open strings in an external constant magnetic field $H$.  For
an
(infinite) sequence of critical values of $H$ an increasing number of (highest
spin component) states lying on the first Regge trajectory becomes tachyonic.
In the limit of infinite $H$ all these states are tachyons (with a common
tachyonic mass) both in the case of the bosonic string and for the Neveu-
Schwarz sector
of the fermionic string. This result generalizes to extended object the same
instability
which occurs in ordinary non-Abelian gauge theories.  The Ramond states have
always positive square masses as is the case for ordinary QED.  The weak field
limit of the mass spectrum is the same as for a field theory with gyromagnetic
ratio $g_S=2$ for all charged spin states.  This behavior suggests a phase
transition of the string as it has been argued
for the ordinary electroweak theory.
\ve

It is well known that the QED vacuum in an environment of a constant magnetic
field $H$ is stable, whatever the value of $H$ is.  This is classically due to
the value of the gyromagnetic ratio $g_{1/2}=2$ for the electron and,
is also true at the quantum level when $g_{1/2}-2 \not=0$, provided the
problem is
correctly handled, as shown by Schwinger long ago.$^{1)}$
This is in sharp contrast
with the case of an environment produced by a constant electric field, where
the QED vacuum suffers from an instability and pair production occurs at a
calculable rate, as computed by Schwinger in 1951.$^{2)}$

The situation drastically changes when electrically charged higher spin
particles are considered.  Here the energy spectrum in a constant magnetic
field reads:$^{1)}$

$$E_n^2=(2n+1)eH-g_SeH\cdot S+m_s^2,\eqno(1)$$

\no where $n$ is the Landau level and $g_S$ is the gyromagnetic ratio of the
particle with spin $s$, charge
$e$ and mass $m_s^2$.

We see that $E_n^2>0$ provided $g_SS\leq 1$.  The value $g_S={1 \over S}$ is
precisely the prescription of ''minimal coupling" for higher spin
particles.$^{3)}$
However this prescription does not hold in consistent theories
coupling higher spin
states.  Indeed we already know that in the standard model, for $W^\pm$
bosons, $g_W=2$ and in generalized gauge theories,$^{4)}$
such as string theories,
$g_S=2$ for all excited string states.$^{4,5)}$  The authors\footnote{*}{The
special occurrence of the value $g_S=2$ for particles of arbitrary spin was
precisely discussed, in different contexts, by V. Bargmann, I. Michel, V.
Telegdi$^{6)}$ and S. Weinberg.$^{7)}$}
of Refs. [8.9] pointed out
that the non-minimal e.m. coupling of $W$ bosons, implied by the non-Abelian
gauge structure, causes an instability of the electroweak vacuum whenever $H$
reaches the critical value$^{9)}$

$$H_{crit}=m_W^2/e\eqno(2)$$

\no as is evident from Eq. (1).  For that value a phase transition may occur,
characterized by $W$-condensation.  Eventually a restoration of the electroweak
symmetry may also take place.$^{9)}$

It is of interest nowadays to consider similar calculations ins tring theory
which provides a model for consistent electromagnetic interactions of particles
of arbitrary spin.  In particular in open string theory, with two charges
$e_1,e_2$ at the string end points, the electromagnetic interaction of all
string states in a constant electromagnetic field strength background can be
handled exactly, i.e. to all orders in the string tension
$\alpha^\prime$.$^{10)}$

This treatment yields a unique set of non-minimal e.m. couplings for the
charged
particle states associated with the string excitations and one may consider
whether the string vacuum is stable in an environment of a given e.m. field
configuration.

The answer to this question was recently given in the case of a constant
electric field.$^{11)}$  It was found that the string vacuum
is unstable against pair
creation with an exactly calculable rate.  This rate reproduces the Schwinger
formula for QED when $eE\alpha^\prime \ll 1$ ($e=e_1+e_2$) but deviates from
the
field theory result when $eE \sim O(1/\alpha^\prime)$, due to the non-minimal
e.m. couplings of the charged string states.

The purpose of this letter is to consider the equally intriguing situation of a
constant magnetic field $H$ where an instability is expected because of Eq.
(1), due to the fact that $g_S=2$ for the higher spin states.  This fact gives
further evidence that string theory is a kind of generalized gauge theory, in
which the string tension parameter $\alpha^\prime$ plays the role of an order
parameter for the breaking of a huge gauge symmetry governing the unbroken
phase.

Since string corrections must be small for $eH\alpha^\prime \ll 1$, Eq. (1)
certainly holds for the spin 1 massless gauge bosons, indicating that the same
phenomenon of instability occurs as in ordinary gauge field theories.
This result was pointed out in Ref. 10 for the open bosonic
string.\footnote{*}{Here we have in mind a string in a constant electromagnetic
(rather than chromomagnetic field so the instability will eventually occur for
some central value of $H$, related to the small $(M_W^2\alpha^\prime \ll 1)$
mass of the $W$ bosons.}
Moreinteresting is to see whether the massive string states, with masses
$O(1/\alpha^\prime)$, become tachyonic for some critical value of $H$.

The answer to this question can be obtained by looking at the analog of Eq. (1)
in open string theory.

We will consider both the cases of bosonic and fermionic open strings.  In the
case of the bosonic string the relevant formula was given in Ref. 10 and
reads

$$\eqalign{\alpha^\prime E^2&=(2b_0^+b_0+1){\epsilon \over 2}-
    {1 \over 2} \epsilon^2\cr
  &-\epsilon \sum_{n=1}^\infty (a_n^+a_n-b_n^+b_n)+L_{free}\cr}\eqno(3)$$

\no where

$$L_{free}=\sum_{n=1}^{\infty} n(a_n^+a_n+b_n^+b_n)-1+L_o^\perp.\eqno(4)$$

\no Here the oscillators $a_n,b_n$, obeying canonical commutation relations,
correspond to the string coordinate $x^1+ix^2$, where $H_{12}\not=0$ (all other
components vanish).  $L_o^\perp$ denotes the Virasoro operator of the
transverse
coordinates with oscillators $\alpha_n^\perp$.

The dimensionless quantity $\epsilon$ reads

$$\epsilon={1 \over \pi}|\arctan 2\alpha^\prime e_1H\pi+
  \arctan 2\alpha^\prime e_2 H\pi|.\eqno(5)$$

\no Notice that for $\alpha^\prime e_iH \ll 1$

$$\epsilon \sim 2\alpha^\prime (e_1+e_2)H,\eqno(6)$$

\no while for $\alpha^\prime e_iH \gg 1, \,\, \epsilon\rightarrow 1$.

To make contact between Eqs. (3) and (1) one identifies $b_o^+b_o$ in Eq. (3)
with the Landau level $n$ in Eq. (1) and notices that

$$\epsilon \sum_{n=1}^\infty (a_n^+a_n-b_n^+b_n)=\epsilon S\eqno(7)$$

\no ($S$ is the spin component along the $H$ direction) and

$$L_{free}=M_S^2\alpha^\prime.$$

\no Formula (3) does reduce to Eq. (1) in the weak field limit $\alpha^\prime
eH\ll 1$, with $g=2$ for all spin states.

Let us investigate if massive states can become tachyonic.  By direct
inspection of Eq. (3) one finds that only states with $b_n^+b_n=0\,\,(n\geq
0),\,\,a_n^+a_n=0\,\,(n>0)$
can become tachyonic.  In this case we are considering states belonging to the
first (parent) Regge trajectory and Eq. (5) becomes

$$\alpha^\prime E^2={\epsilon \over 2}-{\epsilon^2 \over 2}+
   (1-\epsilon)\tilde{n}-1,\quad \tilde{n}=a_1^+a_1.\eqno(8)$$

Tachyonic states appear for

$$\tilde{n}<{1+{\epsilon \over 2}(\epsilon-1) \over
  1-\epsilon} \qquad (\tilde{n}>1,\,\,\forall\epsilon).\eqno(9)$$

\no It is important to notice that as $\alpha^\prime eH\rightarrow\infty\quad
(\epsilon\rightarrow 1)$ more and more states become tachyonic.  In the limit
of infinite magnetic field $\epsilon=1$ all higher helicity states in the first
Regge trajectory are tachyonic with square mass $\alpha^\prime E^2=-1$.  This
result was anticipated in Ref. 10 for the first excited state $\tilde{n}=1$.

We now consider the more interesting situation of superstrings where no
tachyons
are present for $H=0$, unlikes the bosonic case, and spacetime fermions are
also present in the spectrum.  The analog of Eq. (1) takes a different form in
the Neveu-Schwarz (N-S) and Ramond (R) sectors.

Let us first analyze the R-sector, containing all the spacetime fermions and
in particular a light (massless) spin-1/2 (electron-like) state.

A simple extension of the methods of Ref. 10 gives the following expression for
the energy levels

$$\eqalign{\alpha^\prime E^2_R&=(2b_o^+b_o+1){\epsilon \over 2} +
  \epsilon d_o^+d_o-{\epsilon \over 2}\cr
  &-\epsilon \sum^\infty_{n=1}(a_n^+a_n-b_n^+b_n+d_n^+d_n-
  \tilde{d}_n^+\tilde{d}_n)+L^R_{free}.\cr}\eqno(10)$$

\no With respect fo formula (3), we notice that the vacuum shift is zero
instead of ${\epsilon \over 2}(1-\epsilon)-1$ and that ``$d$" fermionic
oscillators are also present.  The zero mode contribution to Eq. (10) agrees
with Eq. (1) for the QED case $(S={1 \over 2},\,\,g=2)$.  Notice that
$L^R_{free}$ can be written as

$$\eqalign{L_{free}^R&=\sum^\infty_{n=1}
  n(a_n^+a_n+b_n^+b_n+d_n^+d_n+\tilde{d}_n^+\tilde{d}_n)\cr
  &+L^\perp_{free}\cr}\eqno(11)$$

\no By substituting this expression in Eq. (10) one obtains a manifestly
positive definite formula.  The lowest mass states are those for which:

$$\eqalign{&d_o^+d_o=0,\quad b_n^+b_n=0,\quad \tilde{d}_n^+\tilde{d}_n=0,\quad
   a_n^+a_n=0\,\,(n\geq 2),\cr
  &d_n^+d_n=0\,\,(n\geq 2).\cr}$$

\no In this case one has

$$\alpha^\prime E_R^2=(1-\epsilon)(a_1^+a_1+d_1^+d_1).\eqno(12)$$

\no In the limit $\alpha^\prime eH\rightarrow\infty\,\,(\epsilon\rightarrow 1)$
all these
states become massless, while the others remain strictly massive.  The absence
of tachyons in the Ramond sector can be understood by the fact that the
massless spin 1/2 state cannot be tachyonic, analogous to the case of QED,
while the positive
square masses of the excited states is a pure stringy phenomenon due to the
fact that $|\epsilon|\leq 1$.

We now turn to the N-S sector.  In this case the analog of Eq. (1) reads

$$\eqalign{\alpha^\prime E^2_{NS}&=(2b_o^+b_o+1){\epsilon \over 2}-\epsilon
  \sum^\infty_{n=1}(a_n^+a_n-b_n^+b_n)\cr
  &-\epsilon \sum^\infty_{n=1/2}(d_n^+d_n-\tilde{d}_n^+\tilde{d}_n)+
  L_{free}^{NS}.\cr}\eqno(13)$$

\no Eq. (13) contains half-integral moded fermionic oscillators.  The ground
state energy is now ${\epsilon \over 2}-{1 \over 2}$.  As usual the tachyon is
eliminated by the GSO projection.

As for the bosonic string, the only states which can become tachyonic belong to
the first Regge trajectory.  They have the form

$$(a_n)^{\tilde{n}}d^+_{1/2}|0\rangle_{NS}\eqno(14)$$

\no where $|0\rangle_{NS}$ is the N-S vacuum.  For these states Eq. (1) becomes

$$\alpha^\prime E^2_{NS}=-{\epsilon \over 2}+(1-\epsilon)\tilde{n}.\eqno(15)$$

\no Eq. (15) gives tachyonic states for

$$\tilde{n}<{\epsilon \over 2(1-\epsilon)}.\eqno(16)$$

\no Again, in the $\alpha^\prime eH\rightarrow\infty$ limit
$(\epsilon\rightarrow 1)$ all (highest helicity component) states on the first
Regge trajectory become tachyonic with square mass

$$\alpha^\prime m^2 = -{1 \over 2}.\eqno(17)$$

\no The tachyonic mass indicates, as in the case of the electroweak theory, an
instability of the superstring vacuum.  For example, the first massive state
becomes tachyonic for $\epsilon=2/3$, that is for $\alpha^\prime
eH\pi=\sqrt{3}\,\,(e_1=e_2=e)$.  This result gets modified when some of the
transverse components are compactified.  For instance, in the case of a torus
compactification with radius $R \gg(\alpha^\prime)^{1/2}$, the first excited
state which becomes tachyonic is

$$d^+_1\big|p={1 \over R}\rangle.$$

\no Here $p$ is the discrete compactified momentum, thus the state has mass
$\alpha^\prime M^2={\alpha^\prime \over R^2}-{1 \over 2}\epsilon$.  This mass
becomes tachyonic at $\epsilon=2\alpha^\prime/R^2$.

The magnetic field instability, in the electroweak theory, gives rise to a
non-zero W and Z
condensate and a new vacuum in which the electroweak symmetry may be restored.
This phenomenon resembles high T phase transitions in point-field theory as
discussed in Ref. 9.  If
a similar situation occurs in string theory, this may indicate that at high
magnetic fields, the string undergoes a phase transition to a new vacuum with
some huge unbroken symmetry, as suggested in Refs. 12,13 or may reach to a
non-critical string vacuum as supported by the analysis of Ref.    An effective
Lagrangian formulation for a string in a constant magnetic field may unravel
the nature of the phase transition and the structure of the new phase.

\vskip.5cm
\ce {\bf Acknowledgements}
\b
\no We would like to thank Julian Schwinger for enlightening conversations.
\vskip.5cm
\ce {\bf References}
\b
\item{1.} J. Schwinger, in {\it Particles, Sources and Fields}, Vol. III
(1989), Advanced Book Classics, Sections 5-6.
\item{2.} J. Schwinger, Phys. Rev. 82 (1951) 664.
\item{3.} L. P. S. Singh and C. R. Hagen, Phys. Rev. D{\bf 9} (1974) 898;
D{\bf 9} (1974) 910.
\item{ } F. J. Belinfante, Phys. Rev. {\bf 92} (1953) 997.
\item{ } K. M. Case, Phys. Rev. {\bf 94} (1954) 1442.
\item{ } C. Fronsdal, Nuovo Cimento Suppl. {\bf 9} (1958) 416.
\item{ } J. Schwinger, in {\it Particles, Sources and Fields}, Addison Wesley,
Reading, MA (1970).
\item{4.} S. Ferrara, M. Porrati, and V. L. Telegdi, Phys. Rev. D{\bf 46}
(1992) 3529.
\item{5.} E. Del Giudice, P. Di Vecchia, and S. Fubini, Ann. Phys. {\bf 70}
(1972) 378;
\item{ } S. Matsuda and T. Saido, Phys. Lett. B{\bf 43} (1973) 123.
\item{ } M. Ademollo {\it et al.}, Nuovo Cimento A{\bf 21} (1976) 77.
\item{6.} V. Bargmann, L. Michel, and V. S. Telegdi, Phys. Rev. Lett. {\bf 2}
(1959) 453.
\item{7.} S. Weinberg, {\it Lectures in Elementary Particles and Quantum Field
Theory}, ed. by S. Deser (MIT Press, Cambridge, MA, 1970), Vol. I.
\item{8.} H. B. Nielsen and P. Olesen, Nucl. Phys. B{\bf 144} (1978) 376.
\item{9.} J. Ambj{\o}rn and P. Olesen, Nucl. Phys. B{\bf 315} (1989) 606;
Nucl. Phys. B{\bf 330} (1990) 193.
\item{10} A. Abouelsaood, C. G. Callan, C. R. Nappi, and S. A. Yost, Nucl.
Phys. B{\bf 280} (1987) 599.
\item{11.} C. Bachas and M. Porrati, Phys. Lett B{\bf 296} (1992) 77.
\item{12.} J. J. Atick and E. Witten, Nucl. Phys. B{\bf 310} (1988) 291.
\item{13.} D. J. Gross, Phys. Rev. Lett. {\bf 60} (1988) 1229.
\item{14.} I. Antoniadis and C. Kounnas, Phys. Lett. B{\bf 261} (1991) 369.

\bye